# Q# , a quantum computation package for the .NET platform

A. S. Tolba, M. Z. Rashad, and M. A. El-Dosuky

Dept. of Computer Science, Faculty of Computers and Information Sciences,
Mansoura University, Mansoura, Egypt.

tolba_1954@yahoo.com,  magdi_12003@yahoo.com, dr_dos_ok@yahoo.com

October  2007

## ABSTRACT

Quantum computing is a promising approach of computation that is based on equations from Quantum Mechanics. A simulator for quantum algorithms must be capable of performing heavy mathematical matrix transforms. The design of the simulator itself takes one of three forms: Quantum Turing Machine, Network Model or circuit model of connected gates or, Quantum Programming Language, yet, some simulators are hybrid.

We studied previous simulators and then we adopt features from three simulators of different implementation languages, different paradigms, and for different platforms. They are Quantum Computing Language (QCL), QUASI, and Quantum Optics Toolbox for Matlab 5.  Our simulator for quantum algorithms takes the form of a package or a programming library for Quantum computing, with a case study showing the ability of using it in the circuit model.

The .NET is a promising platform for computing. VB.NET is an easy, high productive programming language with the full power and functionality provided by the .NET framework. It is highly *readable*, *writeable*, and *flexible* language, compared to another language such as C#.NET in many aspects. We adopted VB.NET although its shortage in built-in mathematical *complex* and *matrix* operations, compared to Matlab.

For implementation,  we first built a mathematical core of matrix operations. Then, we built a quantum core which contains: basic qubits and register operations,  basic 1D, 2D, and 3D quantum gates, and multi-view visualization of the quantum state, then a window for demos to show you how to use and get the most of the package.

## 1. Introducing Quantum Computing

Quantum computing is a promising approach of computation that is based on equations from Quantum Mechanics. A Bit is the basic computational unit of computing. It encodes a 0 or a 1.  A register of *n* bits can store **ANY** *n*-bit number. A *qubit* (quantum *bit*) exists in a *superposition* of states, and encodes the values 1 and 0 simultaneously.  A quantum register of *n* qubits stores **ALL** *n*-bit numbers, i.e. $2^n$ values[1].

**Quantum State**

The quantum state $|\psi\rangle$ represents a qubit if there are $\alpha, \beta \in C$, where C is the set of Complex numbers, such that
$$|\psi\rangle = \alpha|0\rangle + \beta|1\rangle \quad \text{or} \quad |\psi\rangle = \sin\theta\, |0\rangle + \cos\theta\, |1\rangle$$

With $|\alpha|^2 + |\beta|^2 = 1$. $|0\rangle$ and $|1\rangle$ are the computational basis states. Measuring the state $|\psi\rangle$ with respect to $\{|0\rangle, |1\rangle\}$ basis will give $|0\rangle$ with probability $|\alpha|^2$ and $|1\rangle$ with probability $|\beta|^2$. The states $|+\rangle$ and $|-\rangle$ defined as

$$|+\rangle = \frac{1}{\sqrt{2}}|0\rangle + \frac{1}{\sqrt{2}}|1\rangle$$

$$|-\rangle = \frac{1}{\sqrt{2}}|0\rangle - \frac{1}{\sqrt{2}}|1\rangle$$

**Matrix notation**

A 2-level quantum system can store a single qubit state . We will have

$$|0\rangle = \begin{pmatrix} 1 \\ 0 \end{pmatrix} \qquad |1\rangle = \begin{pmatrix} 0 \\ 1 \end{pmatrix} \qquad a|0\rangle + b|1\rangle = \begin{pmatrix} a \\ b \end{pmatrix}$$

Also, we can say that: $|0\rangle = 01$     and $|1\rangle = 10$.   That is $|x\rangle$ = binary (x + 1).

The symbol $|.\rangle$ is called a *ket*, while the symbol $\langle.|$ is called a *bra*. $\langle\psi|$ represents the conjugate transpose of $|\psi\rangle$

$$|\psi\rangle = \begin{pmatrix} z1 \\ \vdots \\ zn \end{pmatrix} \qquad \langle\psi| = \begin{pmatrix} z1 \\ \vdots \\ zn \end{pmatrix}^t = (\overline{z1},...,\overline{zn})$$

where $z1,\ldots, zn \in$ Complex.

Writing $|1\rangle\langle 0| + |0\rangle\langle 1|$ means mapping $|1\rangle$ to $\langle 0|$ and $|0\rangle$ to $\langle 1|$.  Note that $|\psi\rangle\langle\psi| = 1$ .

**Combing Qubits**
Let A and B be quantum systems with state spaces $H_A$ and $H_B$, the state space of the joint quantum system is

$$H_A \otimes H_B$$

For two qubits,

$$|00\rangle = \begin{pmatrix} 1 \\ 0 \end{pmatrix} \otimes \begin{pmatrix} 1 \\ 0 \end{pmatrix} = \begin{pmatrix} 1 \\ 0 \\ 0 \\ 0 \end{pmatrix} \qquad |01\rangle = \begin{pmatrix} 1 \\ 0 \end{pmatrix} \otimes \begin{pmatrix} 0 \\ 1 \end{pmatrix} = \begin{pmatrix} 0 \\ 1 \\ 0 \\ 0 \end{pmatrix}$$

$$|10\rangle = \begin{pmatrix} 0 \\ 1 \end{pmatrix} \otimes \begin{pmatrix} 1 \\ 0 \end{pmatrix} = \begin{pmatrix} 0 \\ 0 \\ 1 \\ 0 \end{pmatrix} \qquad |11\rangle = \begin{pmatrix} 0 \\ 1 \end{pmatrix} \otimes \begin{pmatrix} 0 \\ 1 \end{pmatrix} = \begin{pmatrix} 0 \\ 0 \\ 0 \\ 1 \end{pmatrix}$$

This is made by **tensor product**

$$\begin{pmatrix} a \\ b \end{pmatrix} \otimes \begin{pmatrix} c \\ d \end{pmatrix} = \begin{pmatrix} ac \\ ad \\ bc \\ bd \end{pmatrix}$$

This means that :
- An operation on a single qubit will in general affect all coefficients of the joint state vector.
- A single qubit operation is highly parallel operation.
- Adding a single quantum bit doubles the memory

A note about operations on a Quantum Computer is that apart from measurements (Input/Output Operations) any quantum operation $U$ is *Linear, Length-preserving* (i.e. input vectors and output vectors have the same number of components), *Unitary* (i.e. $UU^t = I$, where $I$ is the identity matrix), *Reversible*, and *Deterministic* .

For detailed discussions on quantum computing and information, you can selectively refer to ([2], [3], [4]).

## 2. Previous Work on Designing a Simulator for Quantum Algorithms.
Any simulator for quantum algorithms must be capable of performing heavy mathematical matrix transforms. The design of the simulator itself takes one of three forms:
- Quantum Turing Machine
- Network Model or circuit model of connected gates
- Quantum Programming Language

yet, some simulators are hybrid.

**Turing machine**
A Turing machine M is a finite device, which performs operations on a paper tape ([3], [5]). This tape is infinite in both directions, and is divided into same-sized squares. At any given time each square of the tape is either blank (B) or contains a single symbol from a fixed finite list of symbols $s_1, s_2 \ldots, s_n$ that form the alphabet of M.
M has a reading head which at any given time scans or reads a single square of the tape.

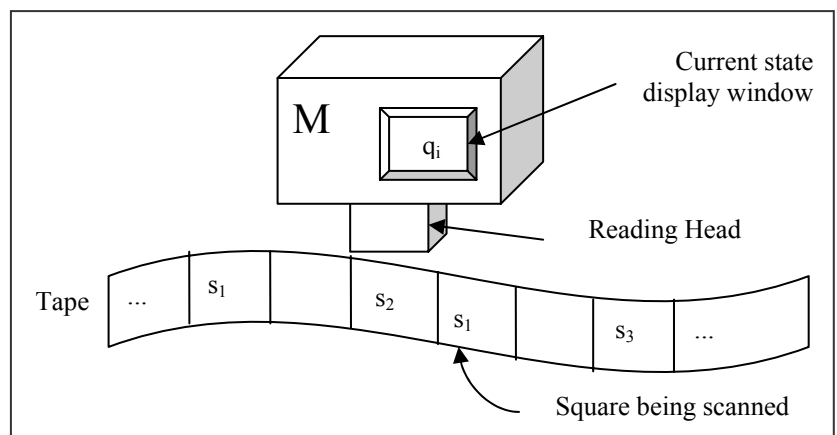

(Figure 1: Turing Machine)

M is capable of three kinds of simple operations: Replacing the symbol in the square being scanned ay another symbol from the alphabet of M, Moving the reading head one square to right, or Moving the reading head one square to left.

The action that M takes at any instant depends on the current state of M and the symbol currently being scanned. This dependence is described in M's specification which consists of a finite set Q of quadruples, each of which takes one of the following forms:

$$q_i\ s_j\ s_k\ q_l$$
$$q_i\ s_j\ R\ q_l$$
$$q_i\ s_j\ L\ q_l$$

$$\begin{pmatrix} 1 \le i, l \le m \\ 0 \le j, k \le n \end{pmatrix}$$

A quadruple $q_i\ s_j\ \alpha\ q_l$ in Q specifies the action to be taken when the state is $q_i$ and scanning the symbol $s_j$, as follows:
1- Operate on the tape thus:
    a)    if $\alpha = s_k$ erase $s_j$, and write $s_k$ n the square being scanned;
    b)    if $\alpha = R$ move the reading head one square to the right:
    c)    if $\alpha = L$ move the reading dean one square to the left:
2- Change into state $q_l$.

**Circuit model of computation**

The circuit model of computation is equivalent to Turing machine model but is nearer to real computers ([2]). The building blocks that can build any circuit are called logic gates. The functionality of logic gates is described in terms of truth table, which specifies all possible configurations of input and the corresponding output. The elementary logic gates with their truth tables are given below, where A and B denote inputs and F denotes output.

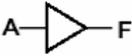
(Figure 2: elementary classical logic gates)

A set of connected gates accomplishing a certain computation is called a circuit. A gate called XOR (*exclusive OR*) and its inverse are given below. Actually, XOR is a common used circuit, so we handle it as if it is an elementary gate.

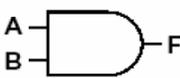
(Figure 3: another two elementary logic gates)

**Note** that: the NOT gate is reversible (as you can guess the input for any given output), while the AND gate is irreversible, so we can say that the AND gate erases information.

Quantum gates are the same as the classical ones, but with maintaining the *Reversibility* and *Length-preserving* by usually outputting extra bits (*ancilla bits*) which usually correspond to a sufficient number of the inputs. For example, the quantum XOR gate takes two inputs, *x* and *y* say, and outputs *x* and the main output that is **x ⊕ y**.

$$|x, y\rangle \rightarrow |x, x \oplus y\rangle$$

Hence, any classical gate can be modified to be used as a quantum gate.

When studying quantum gates we usually classify them according to the number of qubits the gate operates on into 1D (1 dimensional), 2D, and 3D. The common 1D quantum gates are:

$$Identity = \begin{pmatrix} 1 & 0 \\ 0 & 1 \end{pmatrix} \quad NOT = \begin{pmatrix} 0 & 1 \\ 1 & 0 \end{pmatrix} \quad Phase\_Flip = \begin{pmatrix} 1 & 0 \\ 0 & -1 \end{pmatrix} \quad Hadamard = \frac{1}{\sqrt{2}}\begin{pmatrix} 1 & 1 \\ 1 & -1 \end{pmatrix} \quad \sqrt{NOT} = \frac{1}{\sqrt{2}}\begin{pmatrix} 1 & -1 \\ 1 & 1 \end{pmatrix}$$

The common 2D quantum gates are *CNOT* and *SWAP*. The *CNOT* (**Controlled-Not**) is quantum version of the XOR operation and it indicates the interaction between two qubits. This gate has many representations besides

$$|x, y\rangle \rightarrow |x, x \oplus y\rangle$$

| Input/Output transform | Gate | Matrix | Block diagram and function |
|---|---|---|---|
| in    out<br>$\|00\rangle \mapsto \|00\rangle$<br>$\|01\rangle \mapsto \|01\rangle$<br>$\|10\rangle \mapsto \|11\rangle$<br>$\|11\rangle \mapsto \|10\rangle$ | 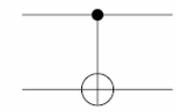<br>first input is the control (black) | $\begin{pmatrix} 1 & . & . & . \\ . & 1 & . & . \\ . & . & . & 1 \\ . & . & 1 & . \end{pmatrix}$ | 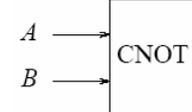<br>If first qubit is set then, apply NOT on the second qubit, else do nothing |

(Figure 4: Different representations of CNOT gate)

The *SWAP* gate can be derived from successive *CNOT* gates.

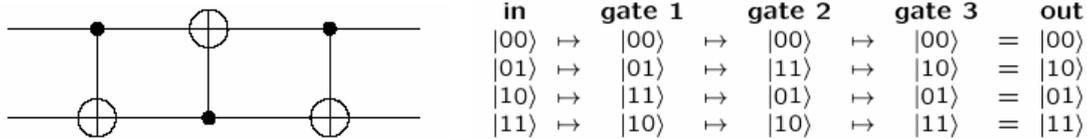

The common 3D quantum gates are *Toffoli* and *Fredkin* which may be considered as the controlled 3D versions of *CNOT* and *SWAP,* respectively. For more about classical and quantum circuit models, you can refer to ([6], [7]) and ([1], [3]) respectively.

**Quantum Programming Language**

Quantum Programming Language is a programming language used to write programs for quantum computers. However, since – at least in the current times – any implementation of a quantum machine has to be controlled by classical device, existing quantum programming languages incorporate classical control structures such as loops (for, while, …) and conditional branching (if, if-else, … ) and allow to operate on classical and quantum data([8], [9]).

The quantum programming language can be either **Imperative** or **functional.** For the imperative quantum programming, we find that the quantum pseudo-code proposed by E. Knill is the first formalized language for description of quantum algorithms([10]). This language was tightly connected with model of quantum machine called Quantum Random Access Machine (QRAM), which is an extension of the classical random access machine but with exploiting quantum resources. Examples for the imperative quantum programming include QCL (Quantum Computer Language), Q Language , and qGCL (Quantum Guarded Command Language).

For the functional quantum programming, we find that during the last few years many quantum programming languages based on functional programming paradigm where proposed, exploiting the advantages of functional programming languages which allow to express algorithms clearly. Examples for the functional quantum programming include Quantum Lambda Calculus, QPL and cQPL.  The first attempt to define quantum lambda calculus was made by Philip Maymin in 1996([11]). In 2003 André van Tonder has defined extension of lambda calculus suitable to prove correctness of quantum programs, with an implementation in Scheme programming language ([12]).

**Well-known Quantum Computing simulators**

Nowadays,  there are many well-known quantum computing simulation implemented in many programming languages such as *Matlab/Octave/Mathematica, Perl/PHP,* and *C/C++/Java.* Some of them are just an editor or a parser and the other may have graphical user interface (GUI). Some of them are standalone and some are just a package or a toolkit to be used in later projects as an extension to the implementing language.

The most known simulators implemented in *Matlab/Octave/Mathematica* are Quack, M-fun for QC Progs, Qubit4matlab , Quantum Octave, QuCalc , QDENSITY, qmatrix , and the Quantum Optics Toolbox for Matlab 5. However the last one is mainly for optic physics, we find it the most comprehensive package in the field and we adopted – besides others – its philosophy and many of  its notations. A remarkable note is that implementing the simulator as a package or a toolkit is common in *Matlab/Octave*/*Mathematica* because of their high extensibility.

The most known simulators implemented in *Perl/PHP* are Quantum::Superpositions, and Quantum::Entanglement. However, the majority of the community prefers *C/C++/Java* especially for applications associated with a master or a doctoral thesis. The most known simulators are Quantum Computer Language (QCL), Open Qubit (library), libquantum, Qubiter(compiler), QCSim , QDD (library), qsims , QGAME(Quantum Gate And Measurement Emulator), QuIDD Pro (Quantum Information Decision Diagram data structure), QuaSi (Graphical algorithm construction), Quantum Algorithm Designer, Quantum Fog(simulator and graphic programming language), Virtual quantum mechanics, jQuantum ( Quantum Computer Simulator ), Quantum Qudit Simulator ,Quantum Computer Emulator (solving Schrodinger equation ), Qdns (GUI, Quantum Designer and Network Simulator), Optical Simulator , Fraunhofer Quantum Computing Simulator (up to 31 qubit)

**Our recommendations and adoptions**

We adopt features from three simulators of different implementation languages, different paradigms, and for different platforms. They are Quantum Computing Language (QCL), QUASI, and Quantum Optics Toolbox for Matlab 5 (let us give it QOTM5 acronym).

**QCL** is the most advanced implemented quantum programming language([13]). It was implemented in C, as a standalone full integrated compiler. Its syntax and data types are similar to those in C. The basic built-in quantum data type is qreg (quantum register), which can be interpreted as an array of qubits. And here is a code snippet in QCL:

```
qureg x1[2];     // 2-qubit quantum register x1
qureg x2[2];     // 2-qubit quantum register x2
H(x1);           // Hadamard operation on x1
H(x2[1]);        // Hadamard operation on the first qubit of the register x1
```

QCL standard library provides standard quantum operators used in quantum algorithms, with the ability to define and use user-defined operators and functions([14]). For example, the following code defines *inverse_about_the_mean* operator used in Grover's algorithm:

```
operator diffuse(qureg q)
{
    H(q);           // Hadamard Transform
    Not(q);         // Invert q
    CPhase(pi,q);   // Rotate if q=1111..
    ! Not(q);       // undo inversion
    ! H(q);         // undo Hadamard Transform
}
```

Within **QOTM5**, the basic data type is a "quantum array" which is a collection of one or more simple "quantum objects"([15]). Each quantum object may represent a vector, operator over some Hilbert space. In the computer, the members of a quantum array object are represented as complex-valued vectors or matrices.

In Matlab, a column vector with *n* components is written as [c1; c2; ...; cn] where the semicolons separate the rows. In order to create a state within QOTM5 and assign it to a variable *psi*, we would write a command such as

>> **psi = qo([0.8;0;0;0.6]);**

The function *qo* constructs a quantum object from a column vector. If then we type *psi* at the prompt, the response is

>> **psi**
**psi = Quantum object, Hilbert space dimensions [ 4 ] by [ 1 ]**
**0.8000**
**0**
**0**
**0.6000**

An object of type *qo* contains the following fields:

| dims | Hilbert space dimensions of each object in the array |
|------|------|
| size | Size of the array, specifying the number of members |
| shape | Shape of each object in the array as a 2D matrix |
| data | Data for the quantum object stored as a "flattened" 2D matrix |

In order to examine these fields, one may enter, for example,

>> **psi.dims**
**[4] [1]**
>> **psi.shape**
**4 1**
>> **psi.data**
**(1,1) 0.8000**
**(4,1) 0.6000**

In order to produce a unit ket in an *N* dimensional Hilbert space, the toolbox function basis(*N, indx*) creates a quantum object with a single one in the component specified by *indx*. Thus we have, for example,

>> **basis(4,2)**
**ans = Quantum object, Hilbert space dimensions [ 4 ] by [ 1 ]**
**0**
**1**
**0**
**0**

QOTM5 defines many basic quantum gates. It also defines many quantum operations such as tensor products.

>> psi1 = qo([0.6; 0.8]);
>> psi2 = qo([0.8; 0.4; 0.2; 0.4]);
>> psi = tensor(psi1,psi2)
psi = Quantum object, Hilbert space dimensions [ 2 4 ] by [ 1 1 ]
0.4800
0.2400
0.1200
0.2400
0.6400
0.3200
0.1600
0.3200

**QuaSi** is a general purpose quantum circuit simulator written at the University of Karlsruhe([16]). It is available on the web as a java-applet. The user is able to build and simulate quantum circuits in a graphical user interface. Key features are that it can simulate up to 20 qubits. It can also realize arbitrarily controlled one-qubit operations. It supports permutation-matrices and can evaluate functions.

The graphical user interface consists of four main windows. The upper left window "Circuit" displays the quantum circuit you want to simulate. The upper right window displays the current system state expressed in the standard basis ($|00..0\rangle$, $|00..1\rangle$, ...$|11..1\rangle$). The lower windows display these amplitudes graphically from left to right. The left window shows the absolute value (length) and the relative phase (direction) of the amplitudes of the basis states, where the phase is indicated by the color and the absolute value by the length of each line. The right window splits the amplitudes into real (blue) and imaginary parts (red) and draws each of them.

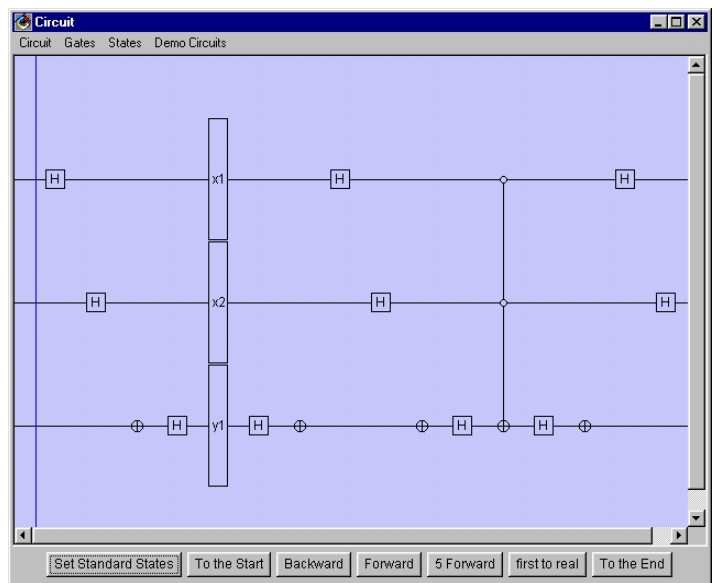
(figure 5: The upper left window "Circuit")

## 3. Microsoft .NET platform

In order to establish the .NET platform, we need to first install the .NET Framework which is Microsoft's latest offering in the world of cross development (developing both desktop and Web applications), interoperability, and, cross-platform development([17]). Recently, Microsoft distributes the framework with Windows operating system. .NET contains functionality that a developer can easily access. This same functionality operates within the confines of standardized data types and naming conventions. The .NET Framework consists of three parts: the Common Language Runtime, the Framework classes, and ASP.NET for web applications.

Now, instead of compiling directly to hardware-specific machine code, the compilation is performed to MSIL (Microsoft Intermediate Language).The syntax of MSIL is similar to machine code, but any MSIL will need to be reinterpreted after it is deployed to the destination machine. This enables a degree of language *interoperability*: Software components from different languages can interact as if they were written in one language. The new Integrated Development Environment (IDE) incorporates some of the best ideas of VB 6.0 and InterDev to make it easier and more intuitive to quickly create applications using a wider variety of development resources. Among the .NET languages there are C#.NET, J#.NET, VB.NET, F#.NET, and many more, all are using and accessing the same resources in the .NET framework.

VB.NET is an easy, high productive programming language with the full power and functionality provided by the .NET framework. Besides its English–like syntax which increases the *readability*, the editor provides many magical auto–completion for all programming constructs which increases the *writability*. Also, VB.NET is a very *flexible* language, compared to another language such as C#.NET in many aspects:
- VB.NET is *case-insensitive*. That is the capital and small cases of a letter are treated the same.
- VB.NET provides both programming styles of *object-oriented* or *structured* programming.
- VB.NET is *customizable*. For example, you can tell the editor to check for variable declarations or not.

For these mentioned reasons and others, we preferred and adopted VB.NET to be the implementation language of the proposed package, although its shortage in built-in mathematical *complex* and *matrix* operations, compared to Matlab.

## 4. The proposed Q#

Initially, we studied previous simulations (esp. QCL, QUASI, and QOTM5) and specified the implementation language to be VB.NET, then we were ready for implementation. First of all, we built a mathematical core of matrix operations. Secondly, we built a quantum core upon the mathematical core. The quantum core contains: basic qubits and register operations, basic 1D, 2D, and 3D quantum gates, and multi-view visualization of the quantum state. A case study utilizing the functionality provided by the package is then implemented. Figure 6 shows the main window for demos to show you how to use and get the most of the package.

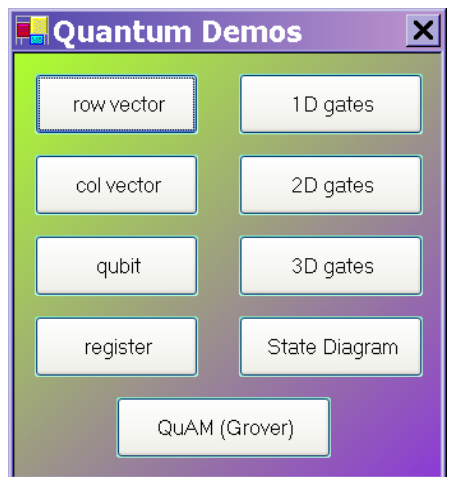

(figure 6: the main window for demos )

**Defining a row vector**

Defining a row vector can be accomplished in many ways. For example, to define the row *r* =[0.8, 0, 0, 0.6], you can write one of the following lines:

        **Dim r As RowVector = row(0.8, 0, 0, 0.6)**
        **Dim r As New RowVector(0.8, 0, 0, 0.6)**
        **Dim r = row(0.8, 0, 0, 0.6)**
        **r = row(0.8, 0, 0, 0.6)**

where the last way assumes either you already declared a variable named *r*, or you state **Option Explicit Off** at the beginning of the file to tell the editor to turn of checking for variable declarations.

It was programmed to make the row takes any number of components as you wish. You can count the number of components in a row *r* by *r.Width* .You can access the component at index *x* by *r(x)*, with x starts from 0. For example:

| r = row(0.8, 0, 0, 0.6) | 0.8, 0, 0, 0.6 |
|---|---|
| r.Width | 4 |
| r(3) | 0.6 |

**Defining a column vector**

Defining a column vector can be accomplished in many ways, with the ease to make the column takes any number of rows. Counting rows of a column *c* is by *c.Height*. Using *tr* on a column performs matrix *transpose*. For example:

| c = col(row(0.8, 0, 0, 0.6), row(1, 0, 1, 0)) | 1, 0, 1, 0<br>0.8, 0, 0, 0.6 |
|---|---|
| c.Height | 2 |
| c(1) | 0.8, 0, 0, 0.6 |
| tr(c) | 1, 0.8<br>0, 0<br>1, 0<br>0, 0.6 |

**Defining qubits and registers**

A qubit can be defined in three different ways: As a column vector, Using the basis function, or Specifying the quantum state, as shown below.

| As a column vector | Hilbert space dimensions [ 2 ] by [ 4 ] |
| --- | --- |
| psi1 = qBit(row(0.8, 0, 0, 0.6), row(0.5, 0.3, 0.2, 0.1)) | 0.8, 0, 0, 0.6<br>0.5, 0.3, 0.2, 0.1 |
| Using the basis function | Hilbert space dimensions [ 2 ] by [ 1 ] |
| psi2 = basis(2, 1) | 1<br>0 |
| Specifying the quantum state, where<br>    d = down<br>    u = up<br><br>psi3 = qState("ddu") | Hilbert space dimensions [ 8 ] by [ 1 ]<br>0<br>1<br>0<br>0<br>0<br>0<br>0<br>0 |

Once you defined qubits, you can use many provided operations such as *scalar product*, *tensor product*, *summing*. For example: **tensor(psi1, psi2)** results in:

    **Hilbert space dimensions [ 4 ] by [ 4 ]**
```
0.8, 0.0, 0.0, 0.6
0.0, 0.0, 0.0, 0.0
0.5, 0.3, 0.2, 0.1
0.0, 0.0, 0.0, 0.0
```

Also you can define a **quantum register** with any number of qubits, of the same size. For example:
    **psi1 = basis(2, 1)**
    **psi2 = basis(2, 2)**
    **r = qReg(psi1, psi2)**
result in
    **Register containing 2 qubits, Hilbert space dimensions [ 4 ] by [ 1 ]**
    0
    1
    0
    0
with ability to access any qubit in such a register. For example: **q = r(0)** yields
    **Hilbert space dimensions [ 2 ] by [ 1 ]**
    1
    0

**Defined quantum gates**
    **1D**: IDENTITY, PHASEFLIP, NOTGate (NOT), SNOT ("Square Root of NOT"), HADAMARD
    **2D**: CNOT= XORGate
    **3D**: FREDKIN, TOFFOLI

For example, calling the *IDENTITY* gate, results in
    **Hilbert space dimensions [ 2 ] by [ 2 ]**
    1, 0
    0, 1

A worth-saying note is that : when programming those gates we did not define them as stored matrices. However, we gave them their quantum operational definition. For example, the *HADAMARD* gate is defined as :

> **Dim** $u$ = qState("u"),   $d$ **As** = qState("d")
>
> **Dim** *result* = sum( tensor(sum($d$, minus($u$)), tr($u$)),  tensor(sum($d$, $u$), tr($d$)))
> **Return** divide(*result*, Math.Sqrt(2))

**Displaying state diagram**

You can display a colorful, highly customizable diagram of one or more quantum state. The default color of the *real* and *imaginary* amplitudes are red and yellow, respectively, with the ability to manually change this sitting. The scales are automatic fixed with the range of displayed data and the size of the window. The state diagram is provided with the ability to change the view into 2D and 3D visualizations such as lines, charts, areas,…etc as preferred.

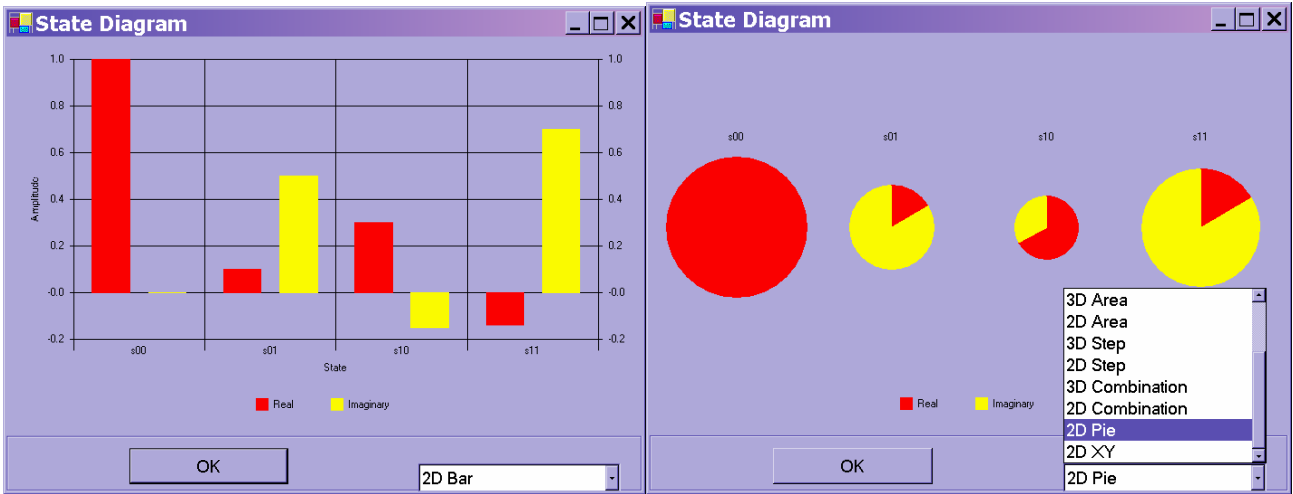
(figure 7: state diagram)

## 5. Case study: Grover Database search

Classically, searching a database of *N* elements for a certain element to specify its index requires *O(N)* comparisons, and $O\left(\frac{N}{2}\right)$ on average. Grover's algorithm achieves the same in $O(\sqrt{N})$ ([18]). Figure 8 shows a sample database.

| index | value |
|-------|-------|
| 1 | **Chicago** |
| 2 | **Cairo** |
| 3 | **Mansoura** |
| ⋮ | ⋮ |

(figure 8: sample search database)

Grover's algorithm operates an *N*-element database, which sometimes is called Quantum Auto-associative Memory (QuAM). The basic idea of Grover's algorithm is to perform a check on all the elements several times and gradually increase the amplitude of the required state. After $O(\sqrt{N})$ iterations, a measurement operation will give the correct answer with probability close to 1.

In the sample database, |x⟩ and |I(x)⟩ correspond to *indices* and *values* respectively. Unitary operators applied are:

$$U_1 = I - 2 \, |w\rangle \langle w|$$
$$U_2 = 2 \, |p_0\rangle \langle p_0| - I$$

where | $p_0$⟩ is the initial state of the database and |$w$⟩, a target state. When a target state |$w$⟩ is 0, then

$$|p_0\rangle = \frac{1}{\sqrt{N}} \sum_{|x\rangle=0}^{N-1} |x\rangle$$

$U_2$ is a reverse transformation on the average, and is called a **Diffusion** operator. $U_1$ is a transformation of selective rotation on a target state, and is called an **Oracle** operator, with the following effect:

$$O|x\rangle = (-1)^{f(x)} |x\rangle$$

If *x* is the required element then the state |*x*⟩ is rotated through π. Where *f(x)* =1 if *x* is the target, and *f(x)* =0 otherwise.

**Tracing Grover's Algorithm**

When it was the time to implement a case study utilizing the package, we decided to re-implement the most of the package, initially in VB.NET, into a web Flash object with ActionScript, and then embed it into a form as if it is built in VB.NET. This weird implementation is for many reasons:
- Challenging our programming skills.
- Better visuality, compared to command-line textual outputs
- Increasing the ability of spreading the package over other platforms other than .NET.
- Execution speed, compared to other web objects such as java applets.

The window for Grover's algorithm is shown in figure 9. At the first sight, the overall design of the window looks similar to that found in adopted Quasi. The window consists of two sky-blue panels with three buttons in between. The upper panel shows the circuit, with a blue indicator to track progress of execution which can be controlled using the *forward* and *backward* buttons. Each step is called a *stage*. The lower panel is for states.

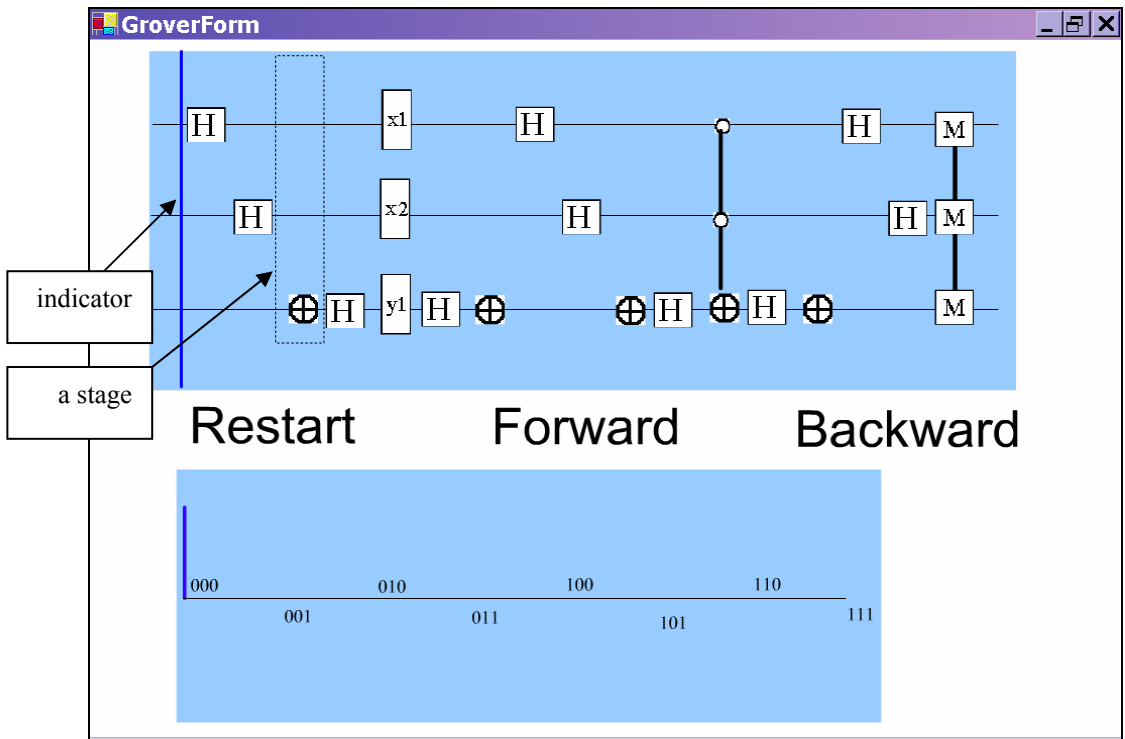
(figure 9: Grover windows)

The algorithm is restricted to the case of searching for an item out of four. We then need 2 (= $\log_2 4$) qubits and a third on for control. To specify the index of the desired item, use *restart* button which shows the following window:

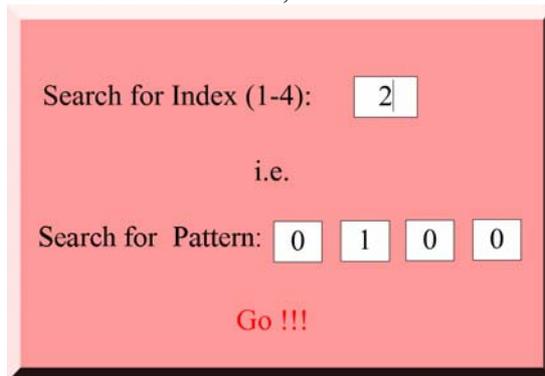
(figure 10: specifying the index)

The initial state is ( |000> 1.0 + 0.0i), and here are some steps of execution:

| | |
|---|---|
| (histogram showing state after stage 1) | **State after stage 1:**<br>\|000> 0.7071067811865475 + 0.0i<br>\|100> 0.7071067811865475 + 0.0i |
| (histogram showing state after stage 2) | **State after stage 2:**<br>\|000> 0.4999999999999999 + 0.0i<br>\|010> 0.4999999999999999 + 0.0i<br>\|100> 0.4999999999999999 + 0.0i<br>\|110> 0.4999999999999999 + 0.0i |

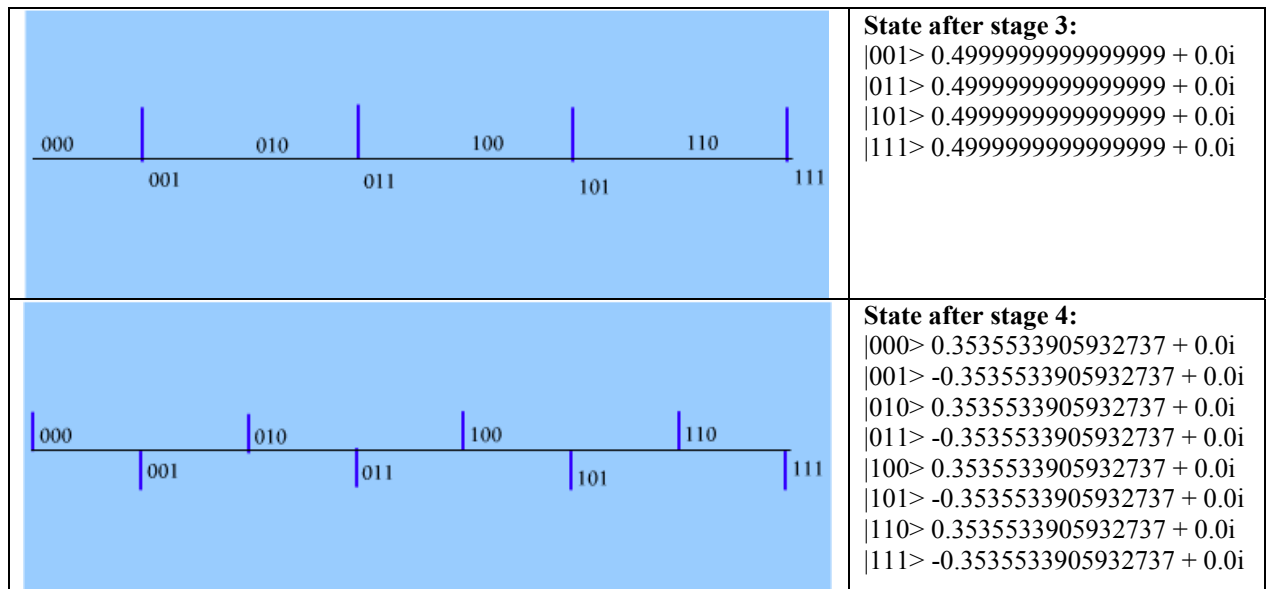

(figure 11: first steps of execution of Grover's algorithm)

we have 16 stages and a final measure (M). A better look shows that we are applying one circuit twice each of 7 stages. We think that gate symbols are self describing as H for *hadamard* and ⊕ for *phase flip*. Stage 5 is the oracle. At the end, we will find the system in the required state.

## 6. Conclusions

Our simulator for quantum algorithms is capable of performing heavy mathematical matrix transforms, and takes the form of a package for a Quantum Programming Language, with a case study showing the ability of using it in the form of Network Model or circuit model of connected gates. We adopt features from three simulators of different implementation languages, different paradigms, and for different platforms. They are Quantum Computing Language (QCL), QUASI, and Quantum Optics Toolbox for Matlab 5.

The .NET is a promising platform for computing. VB.NET is an easy, high productive programming language with the full power and functionality provided by the .NET framework. Besides its English–like syntax which increases the *readability*, the editor provides many magical auto–completion for all programming constructs which increases the *writability*. Also, VB.NET is a very *flexible* language, compared to another language such as C#.NET in many aspects:
- VB.NET is *case-insensitive*. That is the capital and small cases of a letter are treated the same.
- VB.NET provides both programming styles of *object-oriented* or *structured* programming.
- VB.NET is *customizable*. For example, you can tell the editor to check for variable declarations or not.

For these mentioned reasons and others, we preferred and adopted VB.NET to be the implementation language of the proposed package, although its shortage in built-in mathematical *complex* and *matrix* operations, compared to Matlab.

The implementation of the package has many phases. First, we built a mathematical core of matrix operations. Secondly, we built a quantum core upon the mathematical core. The quantum core contains: basic qubits and register operations, basic 1D, 2D, and 3D quantum gates, and multi-view visualization of the quantum state, then a window for demos to show you how to use and get the most of the package.

## 7. Future Work

We should not forget that we have just made a package or a programming library. Although this "steals" the full power of VB.NET, but for the long run it would be better to make it a standalone programming language for sake of independency.

The future will be discussed in three dimensions. The first is *extending the functionality* of the package, a suggestion is to adopt the symbolic representation together with the current numerical. The second dimension is *applying the package* in some possible fields as in image processing. Finally, trying to make it *a standalone full programming language* for sake of independency